\documentclass[12pt,preprint]{aastex}

\usepackage{times}

\def\ion#1#2{#1\,{\sc\romannumeral #2}}

\shorttitle{Spectroscopic Observations of  Fe\,XVIII}
\shortauthors{Teriaca, Warren, \& Curdt}

\begin{document}


\title{Spectroscopic Observations of Fe\,XVIII in Solar Active Regions}

\author{Luca Teriaca\altaffilmark{1}, Harry P. Warren\altaffilmark{2},
  and Werner Curdt\altaffilmark{1}} 
\altaffiltext{1}{Max-Planck-Institut f\"{u}r Sonnensystemforschung,
  Max-Planck-Str. 2, 37191 Katlenburg-Lindau, Germany}
\altaffiltext{2}{Space Science Division, Naval Research Laboratory,
    Washington, DC 20375}


 \begin{abstract}
   The large uncertainties associated with measuring the amount of high
   temperature emission in solar active regions represents a significant
   impediment to making progress on the coronal heating problem. Most
   current observations at temperatures of 3\,MK and above are taken
   with broad band soft X-ray instruments. Such measurements have proven
   difficult to interpret unambiguously. Here we present the first
   spectroscopic observations of the \ion{Fe}{18} 974.86\,\AA\ emission
   line in an on-disk active region taken with then SUMER instrument on
   \textit{SOHO}. \ion{Fe}{18} has a peak formation temperature of 
   7.1\,MK and provides important constraints on the amount of impulsive 
   heating in the corona. 
   Detailed evaluation of the spectra and comparison of the
   SUMER data with soft X-ray images from the XRT on \textit{Hinode}
   confirm that this line is unblended.  We also compare the
   spectroscopic data with observations from the AIA 94\,\AA\ channel on
   \textit{SDO}. The AIA 94\,\AA\ channel also contains \ion{Fe}{18},
   but is blended with emission formed at lower temperatures. We find
   that is possible to remove the contaminating blends and form
   relatively pure \ion{Fe}{18} images that are consistent with the
   spectroscopic observations from SUMER. The observed spectra also
   contain the \ion{Ca}{14} 943.63\,\AA\ line that,
   although a factor 2 to 6 weaker than the \ion{Fe}{18} 974.86\,\AA\
   line, allows us to probe the plasma around 3.5\,MK. The observed ratio
   between the two lines indicates (isothermal approximation) that most of 
   the plasma in the brighter \ion{Fe}{18} active region loops is at 
   temperatures between 3.5 and 4\,MK.
 \end{abstract}



 \section{Introduction}

 The Parker nanoflare model is perhaps the most widely studied
 theory of coronal heating \citep[e.g.][]{parker1988}. In this model
 turbulent motions in the photosphere drive the braiding of magnetic
 field lines which leads to the formation of current sheets and the
 release of energy through magnetic reconnection. A central component
 of this concept is the formation of high-temperature ($\sim$10\,MK)
 emission during the impulsive phase of the reconnection
 \citep[e.g.,][]{cargill1994,klimchuk2001,cargill2004}.

 Considerable observational effort has gone into searching for the high
 temperature emission associated with nanoflares. Recent work with the
 X-Ray Telescope (XRT, \citealt{golub2007}) on \textit{Hinode} by
 \cite{schmelz2009a,schmelz2009b} and \cite{reale2009,2009ApJ...704L..58R}, 
 for example, has
 reported significant emission at temperatures near 10\,MK. Other work
 that has focused on observations with the EUV Imaging Spectrograph
 (EIS, \citealt{culhane2007}) on \textit{Hinode} has found more sharply
 peaked emission measure distributions
 \citep{warren2011,winebarger2011} although significant emission around 6~MK
 is reported by~\cite{2009ApJ...697.1956K}.
 \cite{winebarger2012} has argued
 that even in combination EIS and XRT are capable of setting only
 relatively high upper limits on the active region (AR) emission measure
 above 6\,MK. The highest temperature emission line observed by EIS
 during non-flaring conditions is \ion{Ca}{17}, 
 which has peak formation
 temperature\footnote{Temperature at the peak of the ionization fraction
 from CHIANTI 
 \citep{2009A&A...498..915D}          
 ionization equilibria.} ($T_{f}$) around 5.6\,MK.
 For XRT the broad temperature response makes it difficult to
 identify high temperature, low emission measure plasma if there is
 cooler emission along the line of sight.

 Spectroscopic observations of emission lines formed at high
 temperatures would greatly facilitate the determination of the emission
 measure beyond 6\,MK in solar ARs. Most of the emission
 lines formed at these temperatures, however, lie at soft X-ray
 wavelengths where it is difficult to achieve high spatial
 resolution. Observations with limited spatial resolution taken with the
 Flat Crystal Spectrometer on the \textit{Solar Maximum Mission}
 suggested relatively narrow temperature distributions. Isothermal fits
 to the observed spectra typically yielded temperatures of around 3\,MK
 \citep[e.g.,][]{saba1991,schmelz1996}. The 94\,\AA\ channel of the
 Atmospheric Imaging Assembly \citep[AIA:][]{2012SoPh..275...17L} on the
 \textit{Solar Dynamics Observatory} images \ion{Fe}{18}
 93.94\,\AA, but this wavelength range also includes strong
 emission lines formed at million degree temperatures
 \citep[e.g.,][]{odwyer2010,2012ApJ...745..111T}, making it difficult to 
 use quantitatively. 
 However, structures dominated by \ion{Fe}{18} 93.94\,\AA, emission 
 are observed in AR cores \citep{2012ApJ...750L..10T}. 

 \begin{figure*}[t!]
  \centerline{\includegraphics[clip,angle=90,scale=0.68]{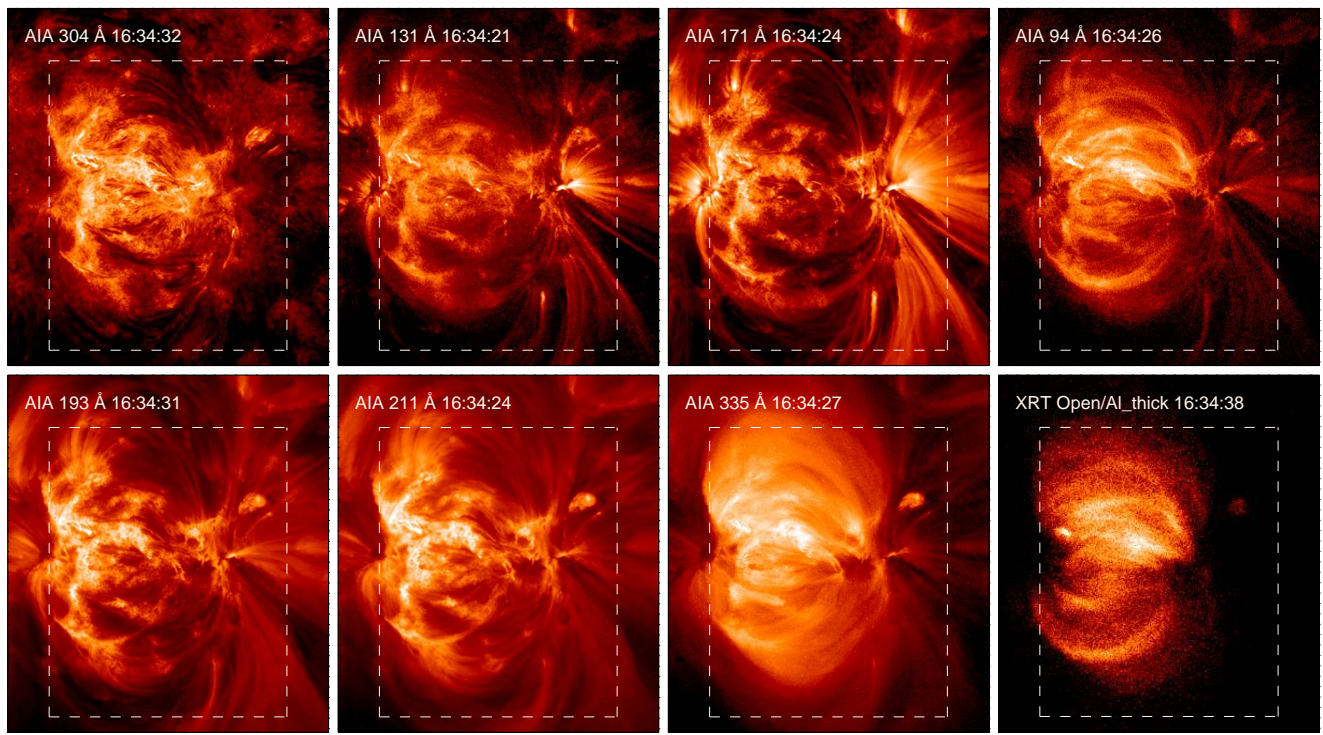}}
  \caption{Observations of active region NOAA-11339 on 2011 November 8
    from AIA and XRT. The dotted line indicates the field of view of
    the SUMER raster. The images have been rotated to
    account for the small \textit{SOHO} roll. The AIA 94\,\AA\ channel
    contains \ion{Fe}{18} 93.94\,\AA\ as well as other unidentified
    lines formed at million degree temperatures. The Al-Thick filter
    is one of XRT's thickest focal plane filters and shows high
    temperature emission. }
 \label{fig:context}
 \end{figure*}
 
 The forbidden transition of \ion{Fe}{18} at 974.86\,\AA\, 
 ($T_{f}=7.1$~MK) is a potentially useful diagnostic for high 
 temperature AR emission. 
 This transition has the highest photon flux per unit emission
 measure of all of the \ion{Fe}{18} lines. It has been observed in
 late-type stars with the \textit{EUV Explorer} and appears to be
 unblended \citep[e.g.,][]{redfield2003}. This line lies within the
 wavelength coverage of the Solar Ultraviolet Measurements of Emitted
 Radiation \citep[SUMER,][]{wilhelm1995} instrument aboard 
 (\textit{SOHO}), but had not been
 observed on the disk previously because the high photon fluxes from the
 nearby \ion{H}{1} Lyman $\gamma$ 972.54\,\AA\ and \ion{C}{3}
 977.07\,\AA\ lines in ARs would have affected the detectors'
 lifetime.

 \begin{figure*}[t!]
  \centerline{\includegraphics[clip,angle=90,scale=0.68]{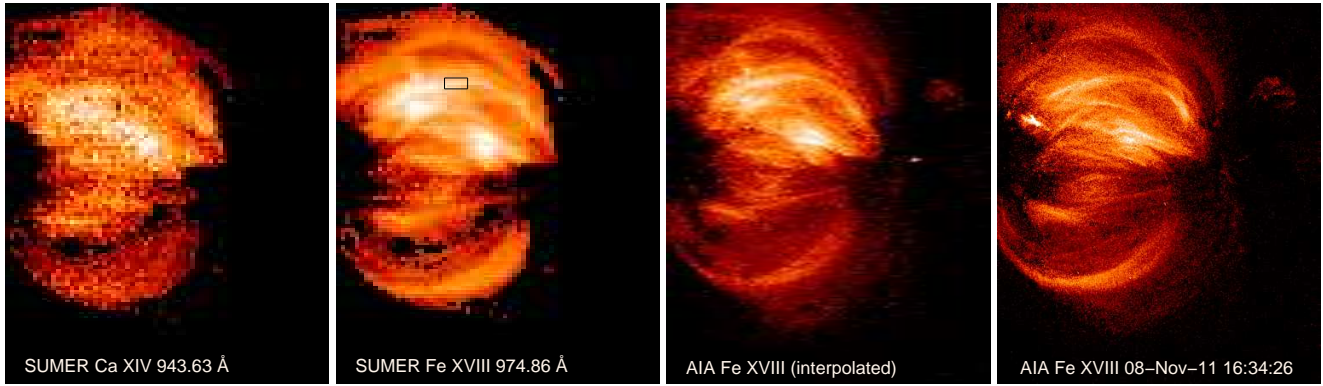}}
  \caption{(\textit{left panels}) SUMER \ion{Ca}{14} and \ion{Fe}{18}
    observations of NOAA-11339 on 2011 November
    8. The black box in the latter panel indicates an area studied in more
    detail later. (\textit{right panels}) The AIA 94\,\AA\ images processed to
    remove the warm emission. The third panel shows the AIA
    \ion{Fe}{18} images interpolated to the times and positions of the
    SUMER slit. The far right panel shows an AIA \ion{Fe}{18} image at
    16:34~UT, the midpoint of the SUMER raster scan.}
 \label{fig:raster}
 \end{figure*}

 In November 2011, a special
 SUMER campaign was organized to observe \ion{Fe}{18} at 974.86\,\AA\ in
 an AR on the disk. Fortuitously, NOAA-11339,
 one of the largest ARs of the current solar cycle, emerged
 during this time and in this paper we present an initial analysis of
 some of the available SUMER, AIA, and XRT data. We confirm that the
 \ion{Fe}{18} at 974.86\,\AA\ shows no detectable blends. Solar emission
 at this wavelength is nearly identical to XRT images in the thickest
 filters. We find that it is possible to remove contaminating
 blends in the AIA 94\,\AA\ channel and form relatively pure
 \ion{Fe}{18} images that are consistent with the spectroscopic
 observations from SUMER. An initial effort at de-blending the AIA
 94\,\AA\ channel has been presented by \cite{reale2011}, but here we
 are able to present a procedure optimized to reproduce the
 spectroscopic observations from SUMER.


 \section{Observations and data analysis}
 
 {\it SUMER}: The observations at the center of this work were taken on
 2011 November 8 between 14:52 and 18:02 UTC by SUMER. 
 In mid-2009, a loss of the gain of the multichannel
 plates was found in the center of the detector's active area. As a
 consequence, to prevent an excessive current leading to a run-away
 effect in the multichannel plates, it was necessary to reduce the high
 voltage. As a result, the overall gain was reduced and in the central,
 KBr-coated, part of the photocathode the gain dropped below detectable
 thresholds. Only the two uncoated areas of the detector (about 200
 pixels each) remain usable. The reduced gain, and some likely
 degradation of mirror reflectivity, has led to an estimated loss in the
 radiometric response of a factor of 2.  In the past, observations of
 very bright areas were forbidden because of the high amount of charge
 they would extract from the detector (reducing the lifetime of the
 multichannel plates) but also because non-linearity effects started to
 affect the data at high count rates by locally depleting the detector
 gain to an extent that becomes unrecoverable above $\approx$ 50
 count/s.  SUMER detectors electronics also produces an electronic
 ghost effect appearing like an echo that becomes visible 2.11\,\AA\ 
 red-ward of bright lines. This last point is
 particularly relevant to observations of the \ion{Fe}{18} 974.86\,\AA\
 line because of the bright \ion{H}{1} line at 972.54\,\AA.
 
 The progressive failure of the detector electronics, and the
 reduced-gain operation, has removed all constraints on the bright
 targets and the SUMER team decided to point for the first time at a
 very active region with high flaring probability: NOAA-11339.

 SUMER is described in detail in the
 instrument paper \citep{wilhelm1995}. Here we just mention that the
 spectra were recorded by imaging the $1''\times300''$ slit with
 detector-B (the only operational after 2006) having an image scale 
 around 960\,\AA\ of $\approx$~44.1\,m\AA/pixel and $1.01''$/pixel,
 respectively.  Data were decompressed, reversed, corrected for response
 inhomogeneities (flat-field), dead-time and local gain effects and for
 the geometrical distortion induced by the read-out electronics.

 The observations discussed here consists of a raster of the target area.
 The study was designed to cover a 300$''\times300''$ area in
 132 steps of 2.2$''$\ (6 elementary steps) for a total of 133 spectra
 (85.5~s exposures). However, the scan mechanism is now loosing up to
 40~\%\ of the commanded elementary steps. Fortunately, the loss
 (verified through encoder readings) is roughly regular
 and the scan could be re-sampled to a regular grid of 70 positions in
 3.04$''$ steps. The effective field of view of the raster is of
 about $211''\times281''$ (pixels at the extremes of the 300$''$ slit
 are lost when correcting for geometrical distortion) and it is shown in
 Figure~\ref{fig:context} overlaid \footnote{Since November 2010 the
   SOHO spacecraft has not maintained the $Z_{0}$ axis of its optical
   reference frame aligned to solar North but to the ecliptic North,
   causing solar images to roll between $\pm 7.25\,^\circ$ that must
   be accounted for.} to AIA and XRT images of NOAA-11339.
 All spectra within $\pm 1.52''$ from each grid location were summed
 together.  The number of summed spectra goes from a minimum of one to a
 maximum of 4.  Total counts are used in line fitting to provide the
 correct error budget (photon noise). 
 Data were further binned over three
 pixels along the slit before fitting the profiles with a single Gaussian 
 plus a linear background leading to the final 
 $3.4''\times 3''$ resolution of the SUMER images.
 
 {\it AIA:} Images from AIA are used to provide context at different
 temperatures (see Figure~\ref{fig:context}). The channel at 94\,\AA\ is
 of particular interest to this study because it should be dominated 
 by \ion{Fe}{18} emission but has also significant contribution from 
 cooler (million degree) lines.
 Thus, a comparison between SUMER and AIA 94\,\AA\ data can help evaluate
 the spectral purity of the latter.

 The contribution of warm (million degree) emission to the AIA
 94\,\AA\ channel can be represented by a linear combination of
 the 171\,\AA\ (\ion{Fe}{9}) and 193\,\AA\ (\ion{Fe}{12}) AIA
 channels and can be evaluated using the
 technique described by 
 \citet{2012arXiv1204.3220W}. 
 The difference between the
 observed 94\,\AA\ AR images and the warm image inferred from
 the 193 and 171\,\AA\ is the AIA \ion{Fe}{18}.

 We have applied this algorithm to each set of 94, 193, and 171\,\AA\
 AIA images during the SUMER observations. For each SUMER intensity we
 have interpolated to the nearest position and time in the AIA
 \ion{Fe}{18} data. The resulting image, which is degraded to the SUMER
 spatial resolution, is shown in Figure~\ref{fig:raster}. For
 comparison, we also show a full resolution AIA \ion{Fe}{18} image taken 
 around  mid-time of the SUMER raster.

\section{Spectral purity of the \ion{Fe}{18} 974.86~\AA\ and 
\ion{Ca}{14} 943.63\,\AA\ lines}

\begin{figure}[t!]
  \centerline{\includegraphics[clip,angle=0,scale=0.65]{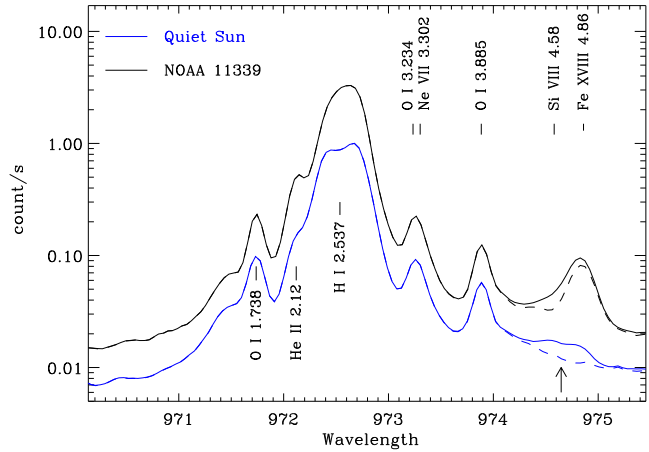}}
  \caption{Average spectrum over NOAA-11339 (solid black) and
    over a quiet Sun area observed on November 10 with the same
    instrumental setting (solid blue). The spectra corrected for the
    electronic ghost (see text) are shown by dashed lines). The arrow 
    indicates
    the location of the electronic ghost induced by the \ion{H}{1}
    972.54\,\AA\ line. Note that the NOAA-11339 spectrum shown is
    averaged over the whole raster including, thus, areas with little or
    no \ion{Fe}{18} emission.}
 \label{fig:blend}
\end{figure}

\begin{figure}[t!]
  \centerline{\includegraphics[clip,angle=0,scale=0.68]{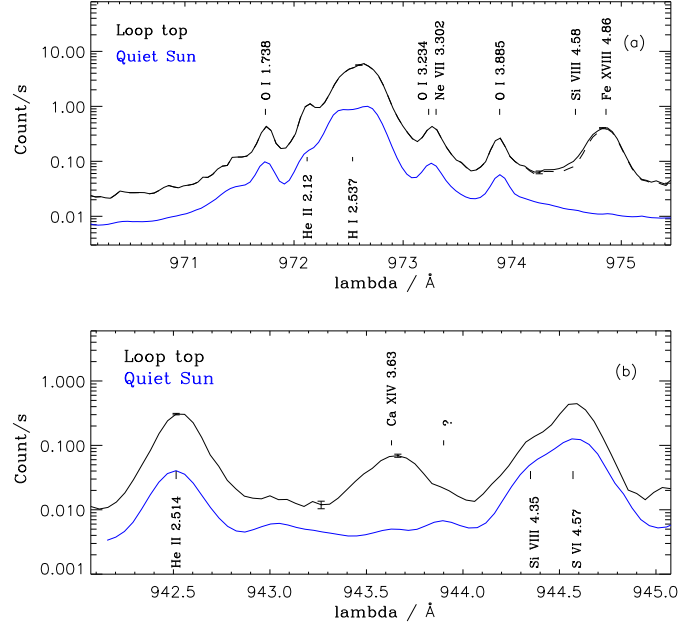}}
  \caption{Spectra obtained averaging over the area marked by a black 
  rectangle in Figure~\ref{fig:raster}.
    {\it (a)}: recorded spectrum of the \ion{Fe}{18} 974.86 line. 
    The spectrum corrected for the
    electronic ghost is also shown by the 
    dashed curve. The corrected
    quiet Sun spectrum of Figure~\ref{fig:blend} is also shown for
    comparison (solid blue).  {\it (b)}: recorded spectrum of the
    \ion{Ca}{14} 943.63\,\AA\ line.  The quiet Sun spectrum from November 
    10 is shown for comparison (solid blue). 
    Representative uncertainties in the AR spectra are also shown.}
 \label{fig:loop}
\end{figure}

{\it Electronic ghost}: It is first necessary to assess blending
with the electronic echo produced by the \ion{H}{1} 972.54\,\AA\ line
around 974.65\,\AA.  Figure~\ref{fig:blend} shows the average recorded
spectrum over NOAA-11339 and over a quiet Sun area observed
on November 10 with the same instrumental setting. Since no
detectable emission from \ion{Fe}{18} is expected from the quiet Sun,
these data are particularly useful for the evaluation of the electronic
blend. Analysis of several datasets acquired over the years with SUMER
shows that, to a first approximation, the ghost can be modeled as 1/200
of the ghosting line shifted 2.11\,\AA\ red-ward \footnote{This is only
  a rough approximation as the shape of the ghost also depends on the
  signal level and there is a distribution of counts all the way from
  the ghosting line to the peak of the ghost. Moreover, the scaling also
  depends on the level of local gain correction applied to the ghosting line.}. 
The result of subtracting such signal from the
observed spectra is shown by dashed lines in Figure~\ref{fig:blend}. It
can easily be seen that no relevant spectral feature is left around
974.86\,\AA\ in the quiet Sun spectrum.  We comment
here that the spectral feature, apparently formed by two un-identified
spectral lines, visible in quiet Sun and coronal hole spectra around
974.6\,\AA\ in \citet{2001A&A...375..591C} 
can be explained by electronic ghosting from the \ion{H}{1} 972.54\,\AA\
line.  In the AR, however, the electronic ghost significantly
affects the blue wing of the \ion{Fe}{18} line and would hamper
velocity measurements unless the \ion{H}{1} line is located outside the
detector. Here we would like to stress that the purpose of this
correction is only to show that no relevant spectral features are
present in quiet Sun spectra after accounting for the electronic ghost
and to provide a first order correction to the \ion{Fe}{18} line
radiance.  The correction is not good enough to safely analyze the line
profile, particularly its blue wing.  No electronic ghost affects the
\ion{Ca}{14} line at 943.63\,\AA\ as there are no bright lines around
941.5~\AA.

{\it Line blends}: The SUMER off-limb coronal atlas
\citep{2004A&A...427.1045C} 
shows a line at 974.58~\AA\ that is identified as due to \ion{Si}{8}
\citep{1999ApJ...512..496K}. 
Other lines from the same $2p{^{2}} 3p$ - $2p{^{2}} 3d$ transitions are
observed at 982.16~\AA, 983.56\,\AA, 988.21\,\AA\ and 994.59\,\AA\
\citep {1997ApJS..113..195F} 
and can be reliably used to check the relevance of the 974.58~\AA\ line.
However, these lines were not acquired when observing NOAA-11339 and we
make a comparison with the \ion{Si}{8} line at 944.34~\AA.  According to
\citep{2004A&A...427.1045C} 
the \ion{Si}{8} 944.34\,\AA\ line is 60 to 100 times stronger than the
974.58\,\AA\ line, while the latter is similar within a factor 2.5 to
the 982.16 and 983.56\,\AA\ lines (a factor 3 in laboratory data
from \citet{1999ApJ...512..496K}.  
 
All the above lines are present in the CHIANTI atomic database
\citep{
2009A&A...498..915D}  
although no experimental data are available for the 974.58\,\AA\ line.
According to CHIANTI the 974.58\,\AA\ line
is hundreds of times weaker than the 982.16 and 983.56\,\AA\ lines but
observations tell us that it should be of a similar intensity, suggesting
problems with the atomic data. On the other 
hand, the 982.16 and 983.56\,\AA\ lines seem to be
in agreement with observations and more than 100
times weaker than the 944.34\,\AA\ line.
Thus, we can expect a
a value between 50 and 100 also for the 944.34/974.58\,\AA\ ratio, as
indeed shown by \citep{2004A&A...427.1045C}.  
Figure~\ref{fig:loop} shows \ion{Fe}{18} 974.86\,\AA\ and \ion{Ca}{14}
943.63\,\AA\ spectra from a selected region at the top of the bright
loop system of NOAA-11339. Figure~\ref{fig:loop}b shows also the
\ion{Si}{8} 944.36\,\AA\ line, telling us that only a negligible signal
is expected in the \ion{Si}{8} 974.58\,\AA\ line.  Similar results are
obtained in different parts of the raster.  In summary, it can be
stated that the \ion{Si}{8} line at 974.58\,\AA\ does not contribute
anything above few percents (if any) to the \ion{Fe}{18} 974.86\,\AA\
line in most of the places where the latter is detectable.

We also note that two lines from \ion{Ar}{7} are listed by NIST at
974.48 and 974.83\,\AA. However, no \ion{Ar}{7} at those wavelengths
have been observed in the experimental work of
\citet{2005EPJD...36...23B}      
and \citet{2007PFR.....2...14K}. 

Figure~\ref{fig:loop}b shows that also the
\ion{Ca}{14} 943.63\,\AA\ line seems substantially unblended. 
In this spectrum, the unidentified line around 943.9~\AA, is 
only about 3~\%, of the \ion{Ca}{14} line and distant enough to 
be negligible when applying a single Gaussian fitting.
On the other hand, the blue 
wing is slightly enhanced (although Gaussian within the uncertainties).
Forcing a third component to the blue wing would results
in a line placed around 943.453 for which we have not found any 
potential candidate in atomic databases and line lists. 
In any case, the difference in the intensity of \ion{Ca}{14} line
from a triple and a single Gaussian fitting is of 8~\%.
Values not larger than 10~\% are also found at other locations 
in the inner part of the AR. 

\section{Summary and discussion}

The \ion{Fe}{18} 974.86\,\AA\ and the \ion{Ca}{14} 943.63\,\AA\ are
unblended spectral lines with formation temperatures of about 7.1 and
3.5~MK, respectively.  As such they allow probing the plasma dynamics at
these temperatures and are critical to establishing the amount of
emission at high temperatures; aspects that are essential to constrain
coronal heating models.

Our observations show \ion{Fe}{18} emission from most of 
the core area of NOAA-11339.
On the other hand, \cite{2012ApJ...750L..10T} find that 
bright emission in the AIA 94\,\AA\ channel and in the EIS \ion{Ca}{17} 
line is observed only in small areas of two other active regions.
The very high activity level of NOAA-11339 is most likely the reason for 
this difference.

\begin{figure}[t!]
  \centerline{\includegraphics[clip,angle=0,scale=0.68]{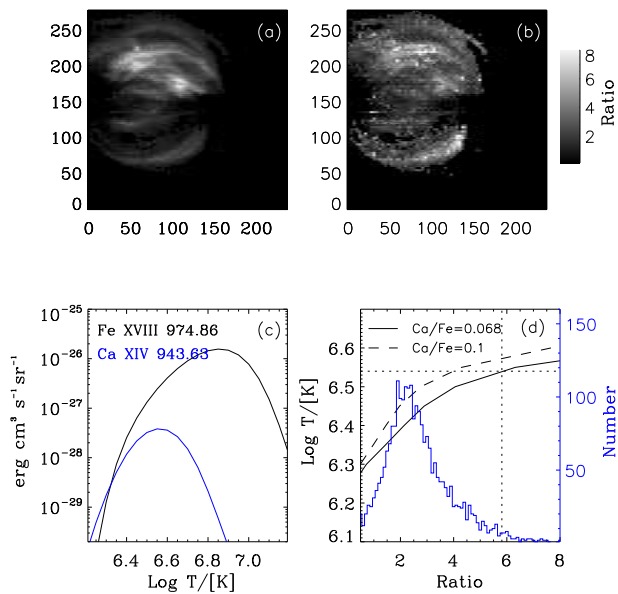}}
  \caption{{\it (a)}: radiance map in \ion{Fe}{18}. 
    {\it (b)}: Observed \ion{Fe}{18} to \ion{Ca}{14} ratio.
    {\it (c)}: contribution functions of the \ion{Fe}{18}
    974.86\,\AA\ and \ion{Ca}{14} 943.63\,\AA\ lines as calculated with
    the CHIANTI database.
    {\it (d)}: \ion{Fe}{18} 974.86\,\AA\ over \ion{Ca}{14} 943.63\,\AA\
    contribution function ratios (black curves). 
    The vertical dotted line indicates
    the ratio measured in the area marked by a black box in
    Figure~\ref{fig:raster} while the horizontal dotted line shows the
    corresponding temperature. The ratio occurrence distribution over 
    NOAA-11339 is represented by the histogram and its relative y-axis (in 
    blue).}
 \label{fig:ratio}
\end{figure}

A quantitative analysis of the amount of emission at high temperatures
would require additional spectral lines (i.e., EIS spectra) to perform a
DEM analysis and a revision of the SUMER absolute radiometric
calibration, both aspects that are beyond the scope of this work.  
However, it seems unlikely that the relative calibration
has changed across the small 943 to 975~\AA\
spectral range, where the radiometric calibration curve is particularly
flat.
This allows some considerations on the \ion{Ca}{14} 943.63 over
\ion{Fe}{18} 974.86\,\AA\ ratio that is, of course, strongly temperature
sensitive.  Since the two ions have similar first ionization potentials
and charge-over-mass ratios, there are no immediate reasons to expect
substantial variations of their abundance ratio with respect to the
value measured in the corona
\citep[0.068,][]{1992ApJS...81..387F} 
and in the photosphere
\citep[0.069,][]{2009ARA&A..47..481A}. 
However, larger values around 0.1 have also been measured in flares
\citep[e.g.,]{1985MNRAS.217..317D}. 
The observed ratio is shown in Figure~\ref{fig:ratio}b.
We plot the contribution functions of the two lines as calculated by
CHIANTI (assuming the CHIANTI ionization equilibria and the coronal
abundances of \cite{1992ApJS...81..387F}) 
in Figure~\ref{fig:ratio}c and their ratio (\ion{Fe}{18} 974.86 over
\ion{Ca}{14} 943.63) in Figure~\ref{fig:ratio}d 
together with the observed ratio distribution. 
In this panel the ratio obtained for Ca/Fe $=0.1$ is also plotted.
The ratio distribution peaks at values around 2.5, corresponding to 
temperatures between 2.5 and 3\,MK, with most of the pixels above these values. 
The brighter areas in 
\ion{Fe}{18} show the largest ratios, with values up to 6. 
As an example, in the area marked by
the black box in  Figure~\ref{fig:raster}, we
measure a ratio of about 5.8 that indicates that most of the
plasma in that region is at temperatures between 3.5 and 4~MK.
We notice that the above value is consistent with the results obtained by
\citet{2012arXiv1204.3220W} 
in a an nearby area of the same loop system, were the above
authors find a DEM (from EIS and AIA data) peaking at log $T$/[K] $=6.6$.

Although the ratio analysis has intrinsic limitations in the very likely case 
of a multi-thermal plasma, the ratios indicate that there is very little plasma 
at temperature larger than 4\,MK (log $T$/[K]$=$6.6).

In summary, the 940 to 980\,\AA\ spectral range has high potential for
the diagnostics of plasmas in the 3.5 to 6.3~MK temperature range
and, particularly, for unprecedented measurements of plasma dynamics at
6.3~MK through spectroscopic analysis of the strong and unblended
\ion{Fe}{18} line at 974.86\,\AA.

The SUMER observations presented here make use of a effective area
0.074~cm$^2$ (without considering the aforementioned further loss of
sensitivity under investigation) that requires long integration, hence
scanning times, that prevent addressing dynamic phenomena. Future
instrumentation with larger effective areas and detectors better capable
of handling high fluxes will certainly provide ground-breaking
observations of the hot plasma above ARs.

The \ion{Fe}{18}~974.86\,\AA\ line will be observed with improved
capabilities by the SPICE spectrograph aboard the Solar Orbiter mission
of ESA.  Moreover, the LEMUR spectrograph \citep{teriaca2011}, proposed
for the Japanese Solar-C mission, will observe these lines in one of its
four simultaneous spectral bands with a 100 times larger effective area
(7.2~cm$^2$) and with higher spatial sampling ($0.28''$), certainly
contributing to a significant step forward in our understanding of the
solar corona.


 \acknowledgments 
      The SUMER project is financially supported by DLR, CNES, NASA and the
      ESA PRODEX programme (Swiss contribution).
      The authors thank U. Sch\"{u}hle for fruitful discussion
      and D. Germerott for his help in acquiring the data. HPW was
      supported by NASA.


\end{document}